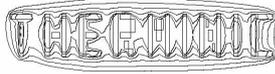

# CONTACTLESS THERMAL CHARACTERIZATION METHOD OF PCB-S USING AN IR SENSOR ARRAY

Gy. Bognár[1], V. Székely[1], M. Rencz[1,2]

[1] Budapest University of Technology, Hungary, <bognar|szekely|rencz>@eet.bme.hu
[2] MicReD, Budapest, Hungary, rencz@micred.com

**ABSTRACT**

In this paper the methodology and the results of a quasi real-time thermal characterization tool and method for the temperature mapping of circuits and boards based on sensing the infrared radiation will be introduced.
With the proposed method the IR radiation-distribution of boards from the close proximity of the sensor card is monitored in quasi real-time. The proposed method is enabling *in situ* IR measurement among operating cards of a system e.g. in a rack, enabling the immediate detection of potential hot spots in the system.

## 1. INTRODUCTION

The elevated temperature encountered in different packaged electronic devices, like digital processors, high power amplifier, high power switches, etc., demands the application of careful temperature-aware design methodologies and the electro-thermal simulations of PCBs.

The results of different electro-thermal simulations and modeling in most of the cases give good approximating results and consider the coupled effects of the real surroundings of these cards and other dissipation elements in an operating system. However the simulation time may take hours, and different systems, different surroundings should be simulated again and again.

In our expectation, by using contactless temperature measurement procedure the heat distribution and the places of high dissipation elements on an operating PCB board (PCI or AGP cards in a rack-house of a PC) can be measured and localized in a dense rack system, where only a thin measuring board needs to be inserted between the operating cards.

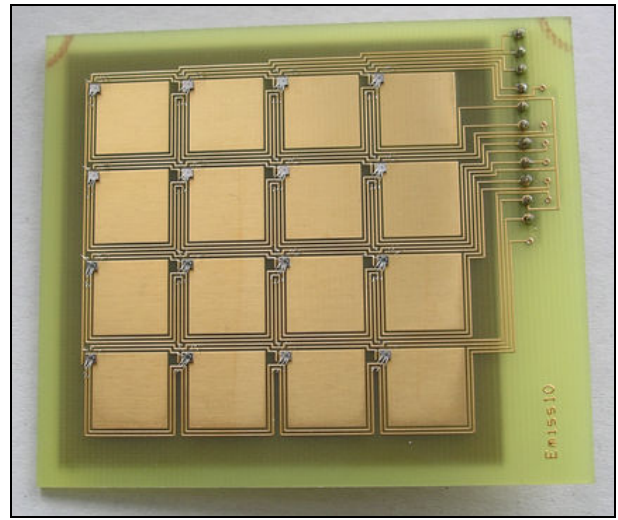

Figure 1. – The sensor card before black painting

There are several contactless temperature measurement methods, but the price and usability determine the application area [1]. The different types of thermographic cameras are not just too expensive but it is impossible to insert them between the cards of a system. So the heat distribution can not be visualized this way. An additional problem is that cooled type IR cameras need liquid gases during the operation [2].

Uncooled IR detectors are mostly based on pyroelectric materials or microbolometer technology [3]. Thanks to the MEMS technology small sensor array can be realized on a silicon die similar to CCD sensors. The only difference is in the sensing method. The CCD sensors are sensitive for the near infrared radiation as well, but they are used mainly for night vision application.

To determine the temperature of a distant object the sensing of the far infrared radiation is needed if the temperature changes of the devices are below 100°C. The main advantage of these sensors is the relatively small (in the 10ms range) time constant. The disadvantage is that additional IR optical elements are needed to get a relatively wide viewing angle. With these additional elements however the all IR sensor system may not be placed between two cards.

By using built-in temperature sensors integrated in the IC-s of the cards only the temperature of the silicon die can be measured, thus the temperature of the packages and the PCB can not be determined, neither the heat distribution along the packages or the board.

In this paper a novel IR measurement method is introduced. In our solution the sensor card will be placed between two system cards, and the heat distribution of one of these PCB panels will be determined and visualized on a display. To check the feasibility of this methodology firstly a small resolution device has been developed for testing purposes.

## 2. THE MEASUREMENT SETUP

Between two cards inserted into two adjacent PCI slots there is about 10mm distance. Our device was realized on a thin (1.55 mm) PCB with a special metallization pattern. In the first experiment a 4x4 matrix was created (Figure 1.). Each square shaped 10mm x 10mm metal "pixel" on the card is aimed for absorbing IR radiation from the neighbourhood to the PCB. The temperature rise due to the absorbed IR radiation is sensed by using bare thermal test-chips, attached directly onto the copper plate of the "pixel" to minimize the thermal resistance between the central point of the pixel and the thermal test-chip. For facilitating good quality ultrasonic bonding the copper, surface on the top side of the measurement card was covered by a 2um gold layer. The applied thermal test chips provide an output frequency depending on the temperature rise of the "pixel". The design of the test chip used in this application is derived from the TMC family of TIMA and MicReD [4]. For minimizing the time constant of each pixel the thermal capacitance had to be decreased by selecting thinner board and thinner metallization layer (35um).

In this application the temperature sensor chips are addressable in the matrix. The signal of the addressed chip with an output frequency proportional to the temperature is available via a simple read-out electronics controlled by an ST7262 microcontroller. The communication between the PC and the electronics is realized by serial link (RS-232) communication.

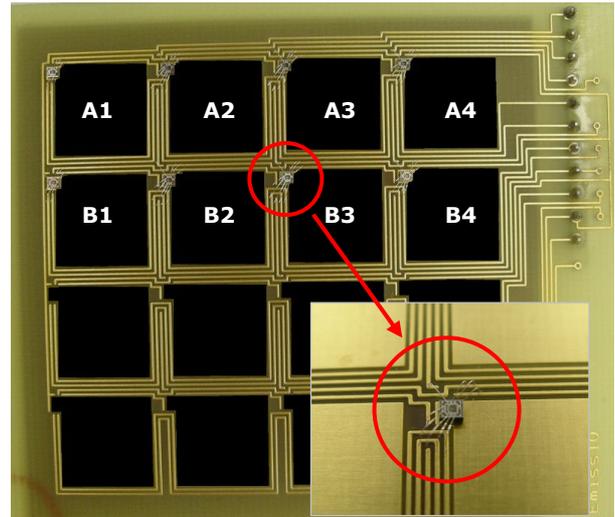

Figure 2. – The temperature sensor dies attached onto the pixels of the sensor card

The wiring of the measurement card is realized only on the top surface of the board without crossing. That is why some space (2…3mm) between the pixels had to be left. The width of the wires (180um) was selected by considering the bonding possibilities and the thickness of the bonding wire. The connector via the communication interface was placed on the right-hand side of the sensor card. For protecting the thermal dies from any damages the chips were covered by a protective polymer drop. The entire board was painted black to enable better absorption of the radiated heat.

In the first experimental realization (Figure 2.) electrically conductive adhesive material with high thermal conductivity was used to attach the chips to the board. In our second experiment, an insulating adhesive was used as die attach, in order to avoid ESD and electrical noise related problems of the temperature sensor chips. The insulating die-attach also realized higher chip-to-board thermal resistance, resulting in increased thermal time-constants of the sensor pixels. On the second boards 8 dies were placed.

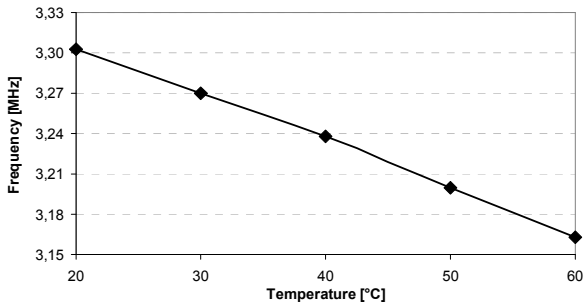

Figure 3. – The temperature vs. measured frequency diagram of the pixel A3

## 3. CALIBRATION AND INITIAL MEASUREMENT

To enable calibration a large, pure copper black-painted plate was placed in front of the sensor card – on which 8 temperature sensing units were placed in a 2×4 matrix form. By using a thermostat the black-painted plate was heated up from room-temperature (21°C) to 60°C degree in 10°C steps.

In every heating step the sensor card was removed until the all black-painted plate reached the adjusted temperature. Then the sensor card was put in front of the black-body, exactly 10 mm far from the plate and in every minute the output frequencies of the sensor chips were recorded as temperature dependent variables.

In Figure 3. the calibration result can be seen for the pixel A3. The approximately 100 kHz difference represents an almost 40°C temperature change.

In the second measurement the black-painted plate was heated up to 50°C and the measurement card was put from 10 mm to 150 mm far from the plate. The obtained temperature vs. distance diagram can be seen in Figure 4. It can be seen that the measured temperature rise shows strong dependence on the distance between the heated and the sensor boards. It means that without calibration process only the place of the highest dissipation elements can be localized but the temperature of the devices attached onto the measured board can not be determined.

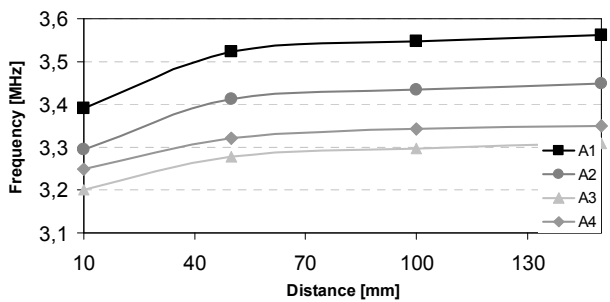

Figure 4. – The distance vs. frequency diagram

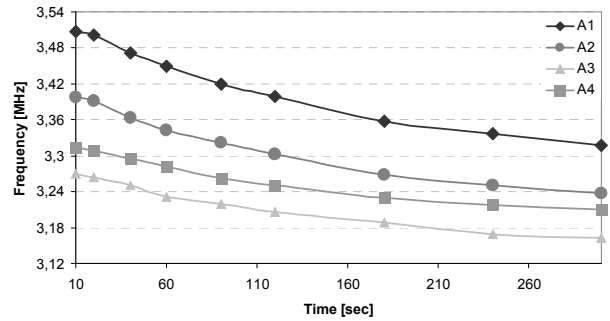

Figure 5. – The time vs. measured frequency function

In Figure 5. the drift of the measure frequency, representing the temperature rise vs. time function can be seen when the calibrator plate was heated up to 60°C in a distant of 10 mm from the sensor card. It can be recognized that the system approaches equilibrium state after 240s. This unfortunately means that fast temperature changes can not be followed by using this method. Because of the physical dimension of the pixels – which means high thermal capacitances as well – we can not reach the 10ms time-constant of the integrated pyroelectric type sensors.

Otherwise if the time constant of the pixels can be decreased significantly (into the range of seconds) it will be appropriate for quasi real-time temperature distribution measurement.

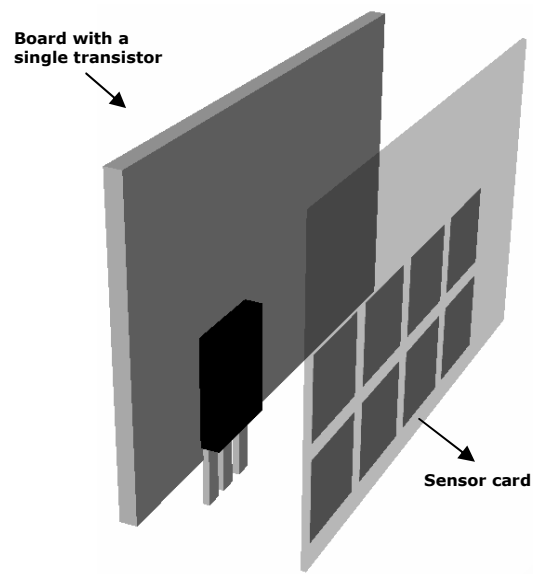

Figure 6. – The measurement setup

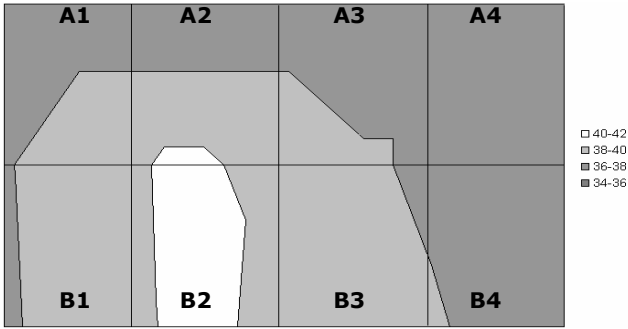

Figure 7. – The obtained heat distribution map

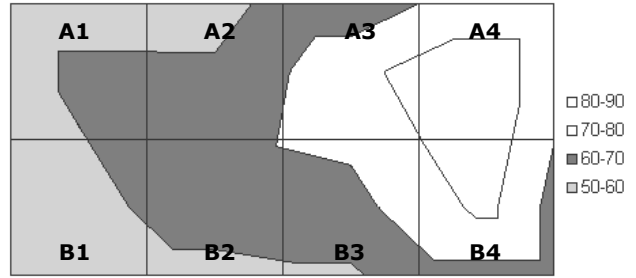

Figure 8. – The measured heat distribution map of the AGP graphics card

Our initial experimental setup already proved the feasibility of this approach.

During our initial measurement a power transistor was used as a heat source. A real board was modeled by using a black painted paper board. The transistor was placed in front of the A2 and B2 pixels, at a 10mm distance from the IR sensing board. A 700mW power step was applied to the transistor for 600 sec. The T3Ster [5] tester was used for driving the transistor and measuring the temperature change of it [7]. It can be seen that the A2 and B2 pixels sensed the biggest temperature changes. The measurement setup can be seen in Figure 6.

The measured temperature changes of the pixels determine exactly the location of the main dissipator element – in this case the power transistor.

Figure 7 shows that the dissipator element was in front of the B2 and A2 pixels during the measurement. The greatest difference between the temperature data of the pixels was 3.5 °C, and the B2 pixel reached the 40.4 °C. It can be seen that at the end of the measurement the IR sensing card warmed up to some extent. Of course the pixels did not reach the same temperature.

## 4. THE MEASUREMENT RESULTS

During the measurement, an AGP graphics card with no built in temperature sensor was investigated during operation. A black-colored heat-sink was attached onto the surface of Graphical Processor Unit (GPU). The aim of our investigation was to find the hot-location i.e. the place of the GPU die under the heat-sink.

The sensor card was placed in front of the graphics card, in a distance of about 10 mm, as shown in Figure 9.

The measured temperature elevation values obtained by the sensor card were uncalibrated, since no calibration could be carried out without using built-in temperature sensor(s) in the GPU. Nevertheless the measurement proved our expectation that it can be used for determine the place of the main dissipation element(s) even under a heat-sink.

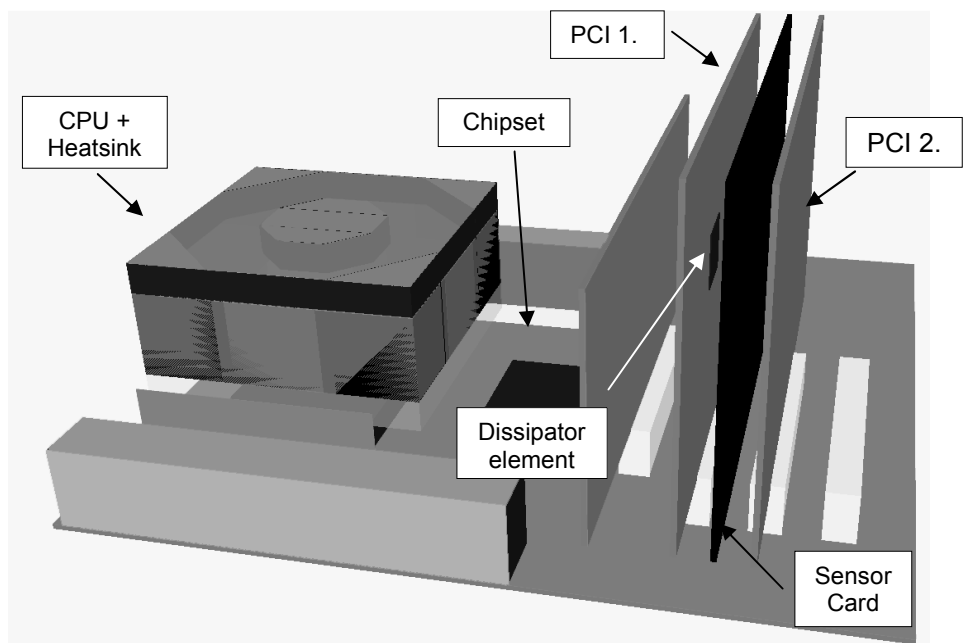

Figure 9. – The measurement setup

### 5. SIMULATION

A couple of simulations have been carried out for cross-verification purposes and to demonstrate the applicability of the sensor card. Steady-state solutions of our measuring system in a complete microATX type desktop rack system considering the heat radiation was done by using the FLOTHERM program [8]. The built-in MicroATX template was completed with the black-painted sensor card placed between the *PCI 1* and *PCI 2* slots. The steady state results show that between the cards inserted into PCI slots the air-flow is minimal. In order to represent the real setup described formerly, a high dissipator devices was placed (10W) onto one of the neighborhood card (PCI 1). In the steady state solution this high dissipator devices reached almost 90°C without any active or passive cooling realizing a hotspot on the surface of a PCI card. The maximum temperature on the surface of the sensor card reached 46°C in the opposite place of the high dissipator element. This result shows correspondence of the results from the measurement, the difference between the simulated and the measured temperature elevations was caused by the heat radiation of the other neighborhood cards.

The temperature distribution on the sensor card inserted into two PCI cards and the place of the highest dissipator elements on the surface of the neighborhood card can be determined. The simulation showed the applicability of the novel IR sensor-card.

In the second simulation an AGP graphics card was investigated. The MicroATX template was completed with the black-painted sensor card, placed in front of the *AGP* slot. Onto the AGP card a high dissipator element (10W) representing the GPU and onto the GPU a small aluminum heat-sink was placed. The temperature of the high dissipator device reached almost 70°C and the maximum "sensed" temperature obtained on the sensor card was 51°C. The difference between the measured and the simulated temperature results is caused because of the different dissipation of the GPU.

The simulation proved the feasibility of determining the exact place of the highest dissipator elements under a heat-sink as well, on the surface of the AGP graphics as it can be seen in Figure 10.

### 6. CONCLUSIONS

We have presented a contact-less method to localize thermal hot-spots. This method can be applicable for sensing the temperature distribution map of a card (PCI, AGP, …) in a dense rack system, where only a thin

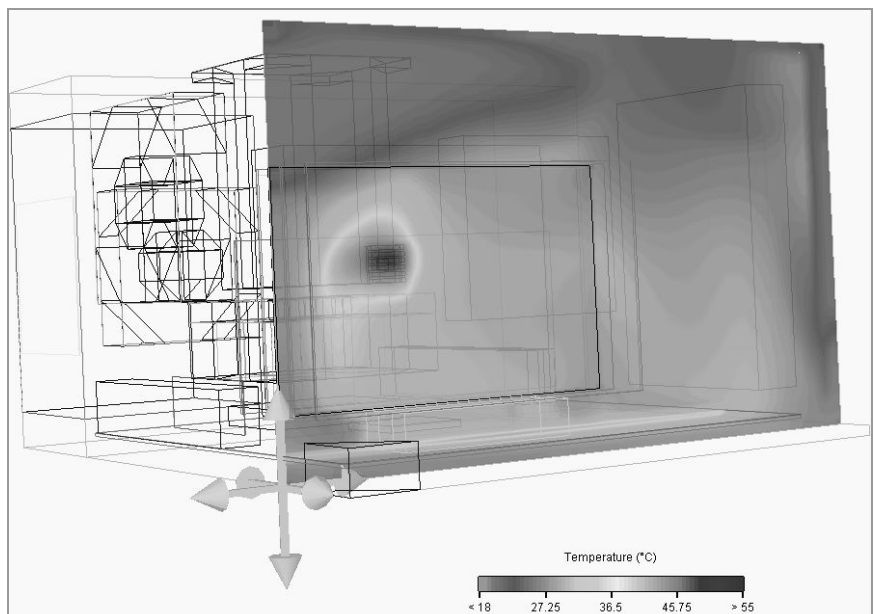

Figure 10. – The FloTherm simulation result of the AGP Graphics Card

measuring board can be inserted between the cards during operation.

The disadvantage of our system is the high thermal resistance between the metal plate and the active area of the temperature sensor integrated circuit. For decreasing the time constant thinner PCB board, maybe back-drilled one under the metal plates and a better thermal conductor adhesive could be applied considering the ground bouncing and substrate noise effects as well.

Application of only one MEMS IR sensor array integrated on one die isn't recommended to substitute the integrated circuits applied in our case, as only a small area of the measured PCB could be visualized. If we substitute our integrated circuits one at a time with several MEMS IR sensors (eg. stand-alone pyrometer or thermopile type sensor), which individually contain also a built in A/D converters (for digital output purposes), the time constant could be dramatically decreased.

## ACKNOWLEDGEMENT

This work was supported by the INFOTHERM NKFP 2/018/2001 Project. The authors would like to thank the help of the Research Institute for Technical Physics and Materials Science of Hungary (MFA).